\documentstyle[twoside,fleqn,espcrc2]{article}

\newcommand{\ttbs}{\char'134}
\newcommand{\AmS}{{\protect\the\textfont2
  A\kern-.1667em\lower.5ex\hbox{M}\kern-.125emS}}

\title{The spin content of the proton in full QCD\thanks{Preprint numbers IFUP-TH
41/97 and IFUM-581/FT.} \thanks{ Partially 
supported by EC Contract CHEX-CT92-0051 and by MURST.}}

\author{B. All\'es\address{Dipartimento di Fisica, Sezione Teorica,
   Universit\`a degli Studi di Milano and INFN, 
   Via Celoria 16, 20133 Milano, Italy.}, G. Boyd\address
	{Center for Computational Physics, University of Tsukuba, 
	Tsukuba, Ibaraki 305, Japan} 
        M. D'Elia\address{Dipartimento di Fisica, Universit\`a di Pisa and 
	INFN, 
        Piazza Torricelli 2, 56126 Pisa, Italy}\thanks{Speaker 
        at the conference.}
        and A. Di Giacomo$^{\rm c}$
        }
       
\begin{document}

\begin{abstract}
We present preliminary results on the proton spin structure
function in full QCD. The measurement has 
been done using 4 flavours of staggered fermions and 
an improved definition of the lattice topological charge
density.
\end{abstract}

\maketitle

\section{INTRODUCTION}

The so--called proton spin crisis 
is still calling for a theoretical 
explanation in the context of QCD.    
The problem may be stated as follows. 
Let $j^5_{\mu}$ be the singlet axial current
\begin{equation}
j^5_{\mu} = \sum_{i = 1}^{N_f} \bar{\psi}_i \gamma_{\mu} \gamma_5 \psi_i.
\end{equation}
Its on--shell nucleon matrix element may be written as
\begin{eqnarray}
\langle \mbox{\boldmath{$p$}}, s | j^5_{\mu} | \mbox{\boldmath{$p$}}', s'  \rangle = \;\;\;\;\;\;\;\;\;\;\;\;\;\;\;\;\;\;\;\;\;\;\;\;\;\;\;\;\;\;\;\;\;\;\;\;\;\;\;\;\;\;\;  \nonumber \\ 
\bar{u}(\mbox{\boldmath{$p$}}, s) \left[ G_1 ( k^2 ) \gamma_{\mu} \gamma_5 -
G_2 ( k^2 ) k_{\mu} \gamma_5 \right] u(\mbox{\boldmath{$p$}}', s'),
\label{form1}
\end{eqnarray}
where $k = p' - p$. In the naive parton model $\Delta \Sigma = G_1 (0)$
can be interpreted as the fraction of the nucleon spin carried by the quarks, 
and so is expected to be close to 1.  
On the contrary, experimental determinations \cite{emc,smc,e143} lead to a 
value $\Delta \Sigma \sim 0.2$, which is also in contrast with the  OZI expectation $\Delta \Sigma \sim 0.7$~\cite{vene1}.
It is the small experimental value of $\Delta \Sigma$ which is 
referred to as the proton spin crisis. 

A numerical simulation on the lattice is the ideal nonperturbative tool 
to obtain an estimate of $\Delta \Sigma$ from first principles.
A direct computation of the matrix element of eq. (\ref{form1}) involves
the evaluation of a disconnected diagram, characterized by a very noisy 
signal \cite{mand1,fuku,liu}. 
We shall adopt a different method \cite{digia1,mand2,alt1}, which 
makes use of the axial anomaly 
\begin{equation}
\partial^{\mu} j^5_{\mu} (x) = - 2 N_f Q(x)
\label{anom}
\end{equation}
$N_f$ is the number of flavours and
\begin{equation}
Q(x) = {g^2 \over 64 \pi^2} \epsilon^{\mu\nu\rho\sigma}
F^a_{\mu\nu}(x) F^a_{\rho\sigma}(x)
\label{qchi}
\end{equation}
is the topological charge density. The on--shell nucleon matrix element 
of $Q(x)$ can be written, using Eq. (\ref{anom}), as
\begin{eqnarray}
\langle \mbox{\boldmath{$p$}}, s | Q(x) | \mbox{\boldmath{$p$}}', s'  
\rangle = \;\;\;\;\;\;\;\;\;\;\;\;\;\;\;\;\;\;\;\;\;\;\;\;\;\;\;\;\;  \nonumber \\ 
= {{m_N} \over {N_f}} A(k^2) \bar{u}(\mbox{\boldmath{$p$}}, s) i \gamma_5 u(\mbox{\boldmath{$p$}}', s'),
\label{form2}
\end{eqnarray}
where $m_N$ is the nucleon mass and $A(k^2) = G_1(k^2) + G_2(k^2) k^2/m_N$, 
whence $A(0) = \Delta \Sigma$. The last relation does not hold in the quenched
approximation, since $G_2(k^2)$ develops a pole at 
$k^2 = 0$ in that case~\cite{mand2}.

Therefore, $\Delta \Sigma$ can be calculated on the lattice by evaluating the
nucleon matrix element of the topological charge density $Q(x)$ in full QCD. 
However, a necessary condition to do this is to have the correct sampling 
of topological modes. Here we will give some preliminary 
results obtained using the 
standard HMC algorithm  with four flavours of dynamical staggered flavours 
and will show that the feasibility of the determination is  affected by 
the bad sampling of topological modes obtained with the HMC 
algorithm~\cite{noi1}. 

\section{THE METHOD}

We have simulated four degenerate flavours of staggered fermions at
$\beta = 5.350$ on a $16^3 \times 24$ lattice, using the standard HMC algorithm 
described in Ref. \cite{Gottlieb87} and the Wilson action for the pure 
gauge sector. A trajectory length $\tau = 0.6$ was used,
with time step $\delta \tau = 0.004$. 
The value $a \cdot m_f = 0.01$ was 
chosen for the fermion mass. 

In order to evaluate the matrix element of Eq. (\ref{form2}) we 
need a lattice regularization $Q_L(x)$ of the topological charge density.
The renormalization properties of $Q_L(x)$ are nontrivial~\cite{haris}:
it mixes with the continuum $Q(x)$, 
$\partial^{\mu} j^5_{\mu} (x)$ as well as with 
$p(x) =  \sum_{i = 1}^{N_f} m_i \bar{\psi}_i  \gamma_5 \psi_i$. 
However, it can be shown that the last two operators can be safely 
neglected in a first approximation~\cite{haris}, so that the following 
relation holds
\begin{equation} 
Q_L = a^4 Z_Q Q + {\cal O}(a^6).
\label{reno}
\end{equation} 

We use the improved smeared operators $Q_L(x)$ proposed
in Ref. \cite{smear}, and in particular the operator defined at the second
smearing level. The value of $Z_Q$ can be obtained 
non--perturbatively using the so--called heating method 
\cite{digia2,noi2}: the short range fluctuations responsible 
for the renormalization are thermalized by applying a few 
updating steps on a discretized classical configuration of charge 1.
The thermalization shows up as a plateau in
$\langle Q_L \rangle$ plotted against the number of heating steps performed, 
as is shown in Fig. 1. $Z_Q$ is then measured on this plateau.
We have obtained $Z_Q = 0.56 \pm 0.04$ with the values of the parameters
used in our simulation.

\begin{figure}
\vspace{3.7cm}
\includegraphics{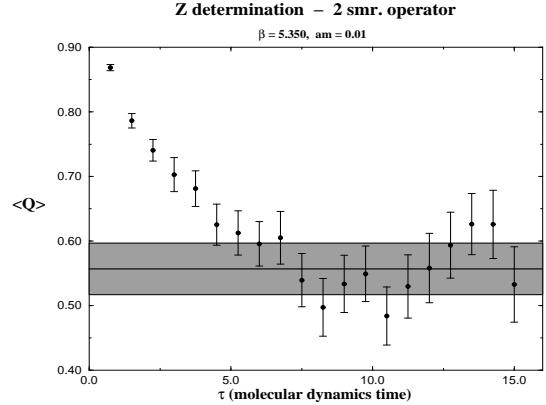} 
\null\vskip 0.3cm
\caption{Determination of $Z_Q$ for the 2--smeared operator. The number of 
heating steps performed is measured in terms of molecular dynamics time.}
\end{figure}

The next step is to compute the correlation function
\begin{equation}
C(t) = \pm { i \over {2 N_f Z_Q | \mbox{\boldmath{$p$}} |}} 
       \langle \bar{B} (\mbox{\boldmath{$0$}},0) P_{\pm} Q_L 
(\mbox{\boldmath{$p$}}, \tau) B( \mbox{\boldmath{$p$}},t)
       \rangle,
\label{correl}
\end{equation}
where 
\begin{equation} 
Q_L (\mbox{\boldmath{$p$}}, \tau) = \sum_{\mbox{\boldmath{$x$}}} 
e^{i \mbox{\boldmath{$p x$}}} Q_L (\mbox{\boldmath{$x$}}, \tau),
\label{fourier}
\end{equation}
and $P_{\pm}$ is the spin projection operator. 
For $\bar B$ we have taken the wall source, after performing a Coulomb 
gauge fixing on the $t = 0$ time slice by means of a steepest--descent 
algorithm. For $B$ we have taken the usual local baryon 
operator~\cite{alt1,alt2}
\begin{equation}
B = \epsilon_{a b c} \chi^a \chi^b \chi^c. 
\label{baryon}
\end{equation}

The following behaviour is expected for $C(t)$~\cite{alt1,alt2}
\begin{eqnarray}
C(t) = 
&&A_+ \;  \Delta \Sigma \; e^{-m_N \tau - E_N (t - \tau)} +  \nonumber \\
&&A_- \; \Delta \Sigma_{\Lambda} \; (-1)^t e^{-m_{\Lambda} \tau - E_{\Lambda} (t - \tau)}, 
\label{fit}
\end{eqnarray}
where $A_{+(-)}$ and $m_{N(\Lambda )}$ can be determined by a fit to the full 
baryon propagator $\langle \bar{B} (0) B(t) \rangle$~\cite{alt1,alt2}. 
$\Delta \Sigma_{\Lambda}$ and $m_{\Lambda}$ are the axial
baryonic charge and mass of the $\Lambda$(1405) particle respectively~\cite{alt2}.

\section{RESULTS AND CONCLUSIONS}

We have collected a sample of 230 configurations, each separated
by 15 HMC trajectories, using 700 hours of a 25 Gflops APE/QUADRICS 
machine.

In Fig. 2 we show the topological charge distribution in our sample. 
It is clear that the correct thermalization for the topological charge 
has not yet been achieved: $\langle Q \rangle \not= 0$ and the distribution 
is not symmetric under $Q \to -Q$. The time history of the 
topological charge is shown in Fig. 3, where the inefficiency of the HMC 
algorithm to change the topological sector is clearly visible.

Despite this deficiency in our ensemble of configurations, we have attempted an 
estimate of $\Delta \Sigma$ and $\Delta 
\Sigma_{\Lambda}$.
From a fit to the full baryon propagator we have obtained
\begin{eqnarray}
&&A_+ = \;\;\,0.152(12), \;\;\; m_N = 0.693(11) \nonumber \\
&&A_- = - 0.192(35), \;\;\; m_{\Lambda} = 0.797(25) 
\label{fitfull}
\end{eqnarray}
We have thereafter measured the correlation function of Eq. (\ref{correl}).
The zero momentum transfer limit has been approximated by using 
$| \mbox{\boldmath{$p$}}| = 2\pi/16$ ($\sim 600$ MeV on our lattice). An average has been 
performed over the three possible spatial directions for $\mbox{\boldmath{$p$}}$ and 
over the two possible polarizations.   

We have observed a poor signal for $C(t)$ and, by combining data
at $\tau = 4$ and $\tau = 20$, we have obtained
\begin{equation}
\Delta \Sigma  \simeq 0.04 \pm 0.04, \;\;\;\;\;
\Delta \Sigma_{\Lambda} \simeq 0.05 \pm 0.05
\label{results}
\end{equation}

The error analysis has been done by a jackknife method. However, the error
quoted in Eq. (\ref{results}) does not take account of the systematic error
coming from the bad sampling of topological modes. By looking at the effective number 
of topological sectors explored during the simulation (Fig. 3),
we roughly estimate the effective error to be up to 4--5 times the one quoted
in Eq. (\ref{results}).  

Therefore our data can only be regarded as preliminary and  
no conclusive results can be produced unless a correct sampling of topological 
modes is obtained. This goal seems to be hardly feasible for the standard 
HMC algorithm: the possibility of achieving it by using different algorithms
is currently under investigation.

\begin{figure}
\vspace{3.7cm}
\includegraphics{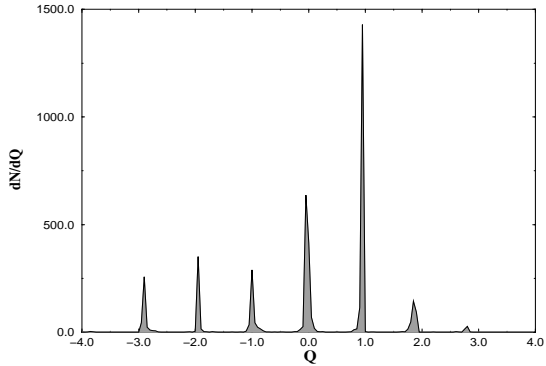} 
\null\vskip 0.3cm
\caption{Topological charge distribution after 30 cooling steps.}
\end{figure}

\begin{figure}
\vspace{3.7cm}
\includegraphics{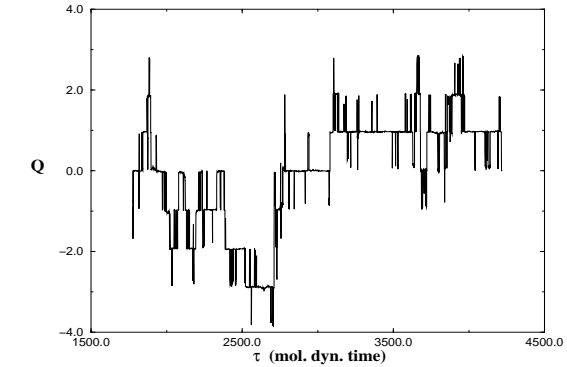} 
\null\vskip 0.3cm
\caption{
  Time history, in units of molecular dynamics time $\tau$, of the topological
  charge $Q$.}     
\end{figure}

\end{document}